# Four-fold truncated double-nested anti-resonant hollow-core fibers with ultralow loss and ultrahigh mode purity

SHOUFEI GAO[1,2,3], HAO CHEN[1,2], YIZHI SUN[1,2,3], YIFAN XIONG[3], ZIJIE YANG[1,2], RUI ZHAO[1,2], WEI DING[1,2,3], YINGYING WANG[1,2,3]*

[1] *Guangdong Provincial Key Laboratory of Optical Fibre Sensing and Communication, Institute of Photonics Technology, Jinan University, Guangzhou 510632, China*
[2] *College of Physics & Optoelectronic Engineering, Jinan University, Guangzhou 510632, China*
[3] *Linfiber Technology (Nantong) Co., Ltd. Jiangsu 226010, China*
*wangyy@jnu.edu.cn*

**Abstract:** Hollow-core fibers (HCFs) are inherently multimode, making it crucial to filter out higher-order modes (HOMs) within the shortest possible fiber length for applications such as high-speed coherent communications and fiber-optic gyroscopes. However, current HCF designs face the challenges of simultaneously achieving ultralow fundamental mode (FM) loss and ultrahigh HOM suppression. In this study, we present a novel four-fold truncated double-nested anti-resonant hollow-core fiber (4T-DNANF) structure that addresses this challenge. Our 4T-DNANF enables greater control over phase-matching between core modes and air modes in the cladding, allowing for minimized FM loss and substantially increased HOM loss. Experimentally, we fabricated several HCFs: one with an FM loss of 0.1 dB/km and an HOM loss of 430 dB/km, and another with an FM loss of 0.13 dB/km with a HOM loss of 6500 dB/km, resulting in a higher-order mode extinction ratio (HOMER) of 50,000—the highest reported to date.

## 1. Introduction

Achieving ultralow loss has been a primary mission within the realm of fiber optics for decades. Over the past half a century, silica-based solid-core fibers have approached the intrinsic Rayleigh scattering loss (RSL) limit of silica glass of 0.14 dB/km [1], leaving little room for further optimization. To push the boundaries of low-loss transmission, researchers have turned their attention to anti-resonant hollow-core fibers (AR-HCF) [2]. These fibers can surpass the fundamental RSL limit associated with solid material [3], as ~99.99% of light is confined within an air core, which has an RSL 4-5 orders of magnitude lower than that of solid glass [4]. Recent advancements in AR-HCF technology have led to significant reductions in attenuation, with losses dropping to a record-low 0.11 dB/km [5]. This progress holds the potential to revolutionize optical fiber-based applications, especially those requiring long fiber lengths and high performance.

With AR-HCF technology achieving this ultimate level of loss performance, researchers are now comparing its optical metrics with those of solid-core fibers across various parameters. In general, AR-HCF offers several advantages, including 31% lower latency, over three orders of magnitude lower nonlinearity, 70% lower chromatic dispersion (compared to standard single mode fibers at 1550 nm), octave-spanning low-loss bandwidth, and significantly higher damage thresholds [2,6,7]. Despite these numerous advantages, one of the primary challenge that remains for AR-HCFs is achieving single-mode operation, which is difficult to resolve due to the inherently multimode nature of HCFs [2,6,8].

From an application standpoint, the key mission of optical fiber communications is to transmit large volumes of data over long distances with higher fidelity. Effectively confining light to a single spatial mode is the main strategy to avoid detrimental multipath interferences and ensure high-fidelity transmission [9]. In fiber optical gyroscopes (FOGs), rotational sensitivity improves with the length of the optical path, but nonlinearities and environmental sensitivities in the fiber glass introduce non-reciprocal errors, which fundamentally degrade measurement accuracy [10]. HCFs have the potential to mitigate these errors but face new challenges arising from their multimode nature [11]. As a strong contender for both applications, HCF must overcome the challenge of simultaneously achieving ultralow loss and ultrahigh mode purity.

From a physics perspective, HCFs are multimode by nature, whether based on photonic bandgap (PBG) or anti-resonant reflection (AR) mechanisms. In PBG-HCFs, both the fundamental mode (FM) and higher-order modes (HOMs) are confined inside the core, as the PBG cladding prohibits light from propagating through it [8]. In AR-HCFs, light is guided via leaky waves [12], with both FM and HOMs satisfying anti-resonant reflection conditions within the core, though HOMs experience greater leakage due to their larger glazing angles [13,14]. This is intrinsically different from solid-core fiber, where total internal reflection guidance mechanism delivers a rigorous definition of single mode operation when the fiber's V-number falls below a certain value (e.g., 2.405 for a rod-shaped cross section) [15].

Several strategies have been proposed to suppress HOMs in AR-HCFs. One approach is to arrange the tubular elements in the cladding asymmetrically, introducing wider gaps that enhance leakage of both FM and HOMs, with HOMs experiencing higher loss [16]. This method is suitable for applications requiring short fiber length, where FM loss is not critical. A more widely adopted approach involves controlling the size of the cladding cavities to allow phase-matching between cladding air modes and the core's HOMs, selectively filtering the latter out [17]. In five-fold double-nested anti-resonant nodeless fibers (5-DNANFs), the middle and small nested tubes are deliberately expanded and contracted, respectively, to create a sufficiently large air cavity between them for phase-matching HOMs[18]. However, this means the effectiveness of the air cavity as an AR element has to be sacrificed in order to strip HOM, creating a dilemma between achieving ultralow loss and robust single-mode operation.

Here, we introduce a novel four-fold truncated DNANF (4T-DNANF) design. We demonstrate that 4T-DNANF can elegantly achieve both ultralow loss and robust single-mode operation, either focusing on one attribute or balancing both simultaneously. Experimentally, with different structural parameters, we demonstrate two combinations: one with an FM loss of 0.1 dB/km and HOM loss of 430 dB/km, and another with an FM loss of 0.13 dB/km and HOM loss of 6500 dB/km. This results in a record-high HOM extinction ratio (HOMER) of 50,000 for HCFs.

## 2. Design Principle

Figure 1(b) illustrates our novel 4T-DNANF design. The purpose of truncating the four large tubes by a 120° segment is not only to reduce the microstructure diameter, as previously done with the five-fold version of the truncated DNANF [19], but more importantly, to prevent the cladding air modes in the voids behind the inter-tube gaps from phase-matching with the core FM, which can occur when they share a similar mode field area. To make a direct comparison, we optimized the structural parameters of both the 4T-DNANF and the five-fold untruncated DNANF (5-DNANF, Fig. 1(a)) to operate at the second-order anti-resonant (AR) band at 1550 nm. We simulated the confinement losses (CL) of the $LP_{01}$ and $LP_{11}$ core modes at 1550 nm and calculated their loss ratio, denoted as HOMER. These simulations were conducted using the finite element method (FEM) in COMSOL Multiphysics with an optimized perfectly matched layer (PML).

The results in Fig.1(d&e) indicate that both the CL and HOMER are strongly influenced by the thicknesses of the first and second cladding air crescents (labeled as $Z_1$ and $Z_2$, respectively) represented by the red and green shaded areas in Fig. 1(b). For the 5-DNANF structure, our simulation in Fig. 1(d) closely aligns the findings reported in Ref. [20], confirming that utilizing the second cladding air crescent (green shaded area in Fig.1(a)) is the most efficient approach in filtering out the $LP_{11}$ core mode while keeping $LP_{01}$ mode loss low. In this configuration, the maximum HOMER achieved is 3500, with the CL of the $LP_{01}$ core mode just below 0.1 dB/km at $Z_2/R_{core}$ of 1.05.

In contrast, the 4T-DNANF structure outperforms the 5-DNANF in terms of both achievable HOMER and CL. Specifically, when $Z_1/R_{core} = 1$, the HOMER in the 4T-DNANF reaches an impressive value of 180,000, with the CL of the $LP_{01}$ core mode remaining below 0.01 dB/km. To investigate the cause of this drastic increase in HOMER, we plotted the effective indices ($n_{eff}$) of relevant modes in Fig. 1(h). At $Z_1/R_{core} \approx 0.72$ ($Z_2/R_{core} \approx 1.06$), phase matching occurs between the $LP_{11}$ core mode and the second cladding air crescent mode, resulting in a HOMER of several thousand—similar to the 5-DNANF case. However, when $Z_1/R_{core} \approx 1$, the $LP_{11}$ core mode phase-matches with the first cladding air crescent mode, boosting the HOMER to 180,000. The proximity of the first cladding air crescent to the core makes it significantly more effective in filtering out HOMs than the second cladding air crescent. This stronger coupling is evident in the intensity distributions of the degenerate modes at the two anti-crossing points, shown in the insets of Fig. 1(h) (log scale).

It is important to note that in the 5-DNANF structure (Fig. 1(d)), although phase-matching between the $LP_{11}$ core mode and the first cladding air crescent mode can also occur at $Z_1/R_{core} \approx 1.05$, the $LP_{01}$ core mode loss increases to an unacceptable 0.5 dB/km, with a HOMER of only 2000 (black dashed circle in Fig. 1(d) and Ref. [20]). This trade-off occurs because enlarging the first cladding air crescent for better HOM suppression requires reducing the size of the middle tube ($Z_2/R_{core} \approx 0.2$). Under this circumstance, when light inside the core attempts to leak away along a certain radial direction (e.g., the yellow dashed arrow in Fig. 1(a)), it may not experience the obstruction of the middle tube, resulting in weak guidance. While this effect is also present in the 4T-DNANF, it is much less pronounced, as the middle tube remains sufficiently large relative to the core size ($Z_2/R_{core} \approx 0.8$, Fig. 1(e)) in the four-fold cladding arrangement. As a result, the CL of the $LP_{01}$ core mode increases only slightly from its minimum value by 0.006 dB/km. The 4T-DNANF structure, therefore, allows for the simultaneous achievement of both high HOMER and low CL.

Next, we examine the minimum achievable CL in both fiber structures, accepting some trade-off in HOMER. For the 5-DNANF, the CL reaches 0.009 dB/km at $Z_2/R_{core} = 0.8$, while the HOMER is only 25. In contrast, the 4T-DNANF achieves a CL of 0.002 dB/km at $Z_1/R_{core} = 0.8$, while maintaining a HOMER of 4000 — higher than the maximum achievable HOMER in the 5-DNANF. Although this level of loss is not yet achieved in experiments, as other loss contributions, including surface scattering loss (SSL), microbending loss (MIBL), and gas absorption loss, all exceed 0.01 dB/km, exploring the minimum CL remains valuable. Not only does it present the fundamental intrinsic loss mechanism, but more importantly, a lower CL provides greater flexibility in adjusting structural parameters for practical applications.

As illustrative examples, we consider two specific scenarios. The first involves reducing the core size to match that of a standard single-mode fiber (SSMF) for intra-data center applications or to minimize the bending loss for gyroscope applications. A smaller core size leads to a larger glazing angle for FM, significantly increasing the leakage loss. It has been predicted that for a DNANF structure, the CL increases as $R_{core}^{-10}$ [6]. To assess the impact of core radius variation on loss performance, we used the minimum loss points from Figs. 1(d) and 1(e) (highlighted by the gray solid circles) as a reference and scale down the structure. The simulation results are shown in the blue curves in Fig. 1(g). For small core radii, the achievable CL of the $LP_{01}$ core mode in 4T-DNANF is 1-2 order of magnitude lower than that in 5-DNANF. At $R_{core} = 7.5$ μm, where the mode field diameter matches that of SSMF, 4T-DNANF exhibits a CL of 10 dB/km, compared to the 360 dB/km CL of 5-DNANF [22], making the former far more suitable for small-core-radius applications.

The second case considers the inter-tube gap size, which is a critical parameter in fiber fabrication. The simulation results in Figs. 1(d) & (e) were based on an optimal inter-tube gap size of approximately 4 μm. However, in practical fiber fabrication, such a small gap requires very high draw tension to prevent the major adjacent tubes from touching. This extreme requirement on drawing tension, in turn, increases the risk of fiber breakage. Relaxing the gap size requirement is an important approach to increase fiber yield. In Fig. 1(g), we fixed all key parameters for the lowest CL in Figs. 1(d) & (e) and only varied the gap size (black curves). If we consider an acceptable CL level of 0.1 dB/km, the permissible gap sizes are 5.4 μm for the 5-DNANF and 7.6 μm for the 4T-DNANF. While 5.4 μm remains challenging, a gap size of 7.6 μm offers greater flexibility in fiber fabrication.

Relaxing the gap size makes our fiber fabrication process, discussed in section 3, more manageable. Accordingly, Fig. 1(f) presents simulation results based on the relaxed structural parameter of a 7 μm gap size, as depicted schematically in Fig. 1(c). In the region where $Z_1/R_{core}$ ranges between 0.4 and 0.8, corresponding to unitizing the first cladding air crescent to filter out HOMs, a CL below 0.04 dB/km with a HOMER of around 10,000 is achievable. When $Z_1/R_{core}$ is between 0.8 and 1, corresponding to unitizing the second cladding air crescent to filter out HOMs, a CL ranging from 0.04 dB/km to 0.15 dB/km with a HOMER of tens of thousands is attainable. In Fig. 1(i), we scan the spectral range from 1300 nm to 1700 nm with two fixed $Z_1/R_{core}$ values of 0.65 and 0.9, representing two strategies: achieving a loss below 0.1 dB/km with an acceptable HOMER under 10,000 or achieving a HOMER above 10,000 with a slightly higher total loss above 0.1 dB/km (allowing room for SSL and MIBL). These two strategies are employed in our fiber fabrication, as discussed in the next section.

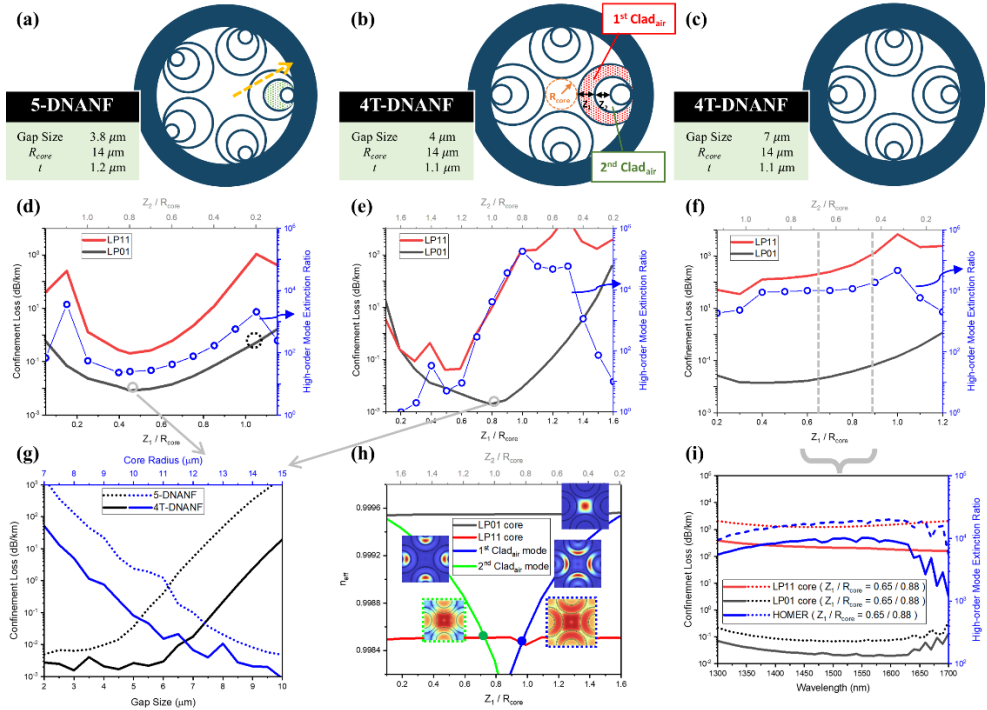

**Fig.1.** Fiber Design and Optical Property Simulations. Cross sections of (a) the 5-DNANF, (b) the 4T-DNANF, and (c) the 4T-DNANF with their structural parameters listed aside. (d,e,f) FEM simulated confinement losses (CL) of the $LP_{01}$ (black), $LP_{11}$ (red) core modes, and HOMER (blue) as a function of $Z_2/R_{core}$ for (a), (b), and (c), respectively, where $Z_2$ is the thickness of the second air layer in the cladding and $R_{core}$ is the core radius. (g) Minimum achievable CL for the structures in (a) and (b) as a function of gap size (black curve) and core radius (blue curve). (h) Effective indices of the core $LP_{01}/LP_{11}$ modes

(black/red), the first cladding air crescent $LP_{01}$ mode (blue), and the second cladding air crescent $LP_{01}$ mode (green), along with their corresponding mode-intensity profiles in linear scale. The plots inside the green and blue dashed squares show the intensity distributions of the core and cladding modes, respectively, on a logarithmic scale. (i) Simulated CL and HOMER as a function of wavelength at two $Z_1/R_{core}$ values of 0.65 and 0.9, using the structural parameters of (c).

## 3. Fiber Fabrication and Characterization

To implement this design strategy, a series of 4T-DNANFs with different $Z_1/R_{core}$ values were fabricated, with their scanning electron microscopy (SEM) images shown in Figs. 2(a-e). These fibers were produced using a modified two-step stack-and-draw method, where large tubes were pre-cut at specific angles with high power lasers before stacking. The key structural parameters are listed in Fig. 2(j). We focus on Fiber #1 and Fiber #5 for detailed analysis of their optical performance.

Fiber #1 has an average $Z_1/R_{core}$ of 0.65, corresponding to the simulation results in Fig.1(i), solid curves designed for a total loss of less than 0.1 dB/km. A total fiber length of 4.2 km is fabricated. Given this relatively short fiber length at this ultralow loss level, the loss value must be determined with great care. Here, three independent loss measurements were conducted with the fiber wound on a standard drum with a diameter of 32 cm. The first measurement involved a cutback loss assessment from 4.2 km to 20 m using a supercontinuum source and an optical spectral analyzer (OSA). The supercontinuum was butt-coupled to 4T-DNANF through a mode field adaptor to minimize the excitation of HOMs. For both long and short lengths, three traces with different output fiber cleaves were acquired to minimize cleaving and OSA coupling errors, with their averaged traces shown in blue and red cuves of Fig. 2(f). The measured loss spectrum, depicted in the black curve in Fig. 2(f), exhibits an attenuation baseline below 0.1 dB/km from 1514 nm to 1600 nm, covering most of the C+L band. The measurement uncertainty is labeled by the shaded area. At 1550 nm, the loss is $0.09 \pm 0.01$ dB/km. The spectral peaks from 1480 to 1510 nm are due to the atmospheric water absorption [21]. Another major absorption peaks in C+L band originates from atmospheric gases, which can hardly be detected at 0.5 nm resolution on the OSA. However, by increasing the resolution to 0.05 nm and repeating the cutback measurement from 1560 to 1620 nm, we can clearly observe loss peaks with frequency spacing of several tens of GHz, resulting in an additional 0.1 dB/km loss at specific wavelengths. Database records indicate that these peaks strongly overlap with the vibrational and rotational absorption lines of $CO_2$ molecules [22]. To eliminate all these peaks, careful purging of the fiber preforms to prevent air from entering the core is necessary, which will be the focus of our next phase of work. The second independent loss measurement was performed at a fixed wavelength of 1550 nm with a stable distributed feedback (DFB) laser and a power meter. Ten data points were recorded at 5-minute intervals, each with a new cleave, before and after the cutback, to ensure minimal error. This resulted in a loss figure of $0.1 \pm 0.01$ dB/km (pink star point in Fig. 2(f)). The third measurement was done using Optical Time Domain Reflectometry (OTDR) at 1550 nm. A comparison with commercially available G.652D an G.654E fibers showed a trace with a lower backscattered intensity but a gentler slope, corresponding to a fitted loss of 0.108 dB/km (Fig. 2(h)). Note that the as-drawn fiber has a sub-atmosphere pressure [23], and since the fiber had been exposed to the atmosphere for some time, air enters the fiber from both ends. The stronger backscattering from the air compared to the glass surfaces means that the steep trace at the beginning and the end actually represents the distribution of gas rather than the fiber loss [24]. Therefore, only the middle section from 1 km to 3.5 km is used to determine the slope. Based on these three sets of measurements, we conclude that fiber #1 achieved a loss of $0.1 \pm 0.02$ dB/km in the C band. It is possible that in certain wavelength regions, the loss has reached even lower level. However, accurately determining such a low loss level requires fiber lengths exceeding 10 km, which will be a focus of future work. Given that the simulated CL reached 0.02 dB/km in Fig. 1(i), we speculate that

SSL and MIBL contribute approximately 0.08 dB/km to the overall loss, and these two loss components could be further reduced with technical advancements. We anticipate that achieving a total loss well below 0.1 dB/km is feasible in the near future.

Fiber #5 has an average $Z_1/R_{core}$ of 0.88, corresponding to the simulation results in Fig. 1(i), dashed curves, designed for very high HOMER. The loss of the FM was measured in the same way as Fiber #1 and shows a loss level of 0.13 ±0.01 dB/km at 1550 nm, as shown in Fig. 2(i). To measure the loss of the HOMs, a combination of cutback and spatially and spectrally ($S^2$) resolved imaging measurement was conducted with a scanning bandwidth of 20 nm from 1540 nm to 1560 nm and a resolution of 0.2 nm [25]. Only 7 m of Fiber #5 was used, as no HOMs could be observed for longer fiber lengths even with offset coupling, which was employed to maximize the intensity of HOMs. The fiber was cut back incrementally to lengths of 6, 5 and 4 m, and at each length, multiple $S^2$ measurements were performed with different cleaves. A linear fit was applied to determine the mode loss. The results, shown in Figs. 2(j) & (k), reveal a $LP_{11}$ mode loss of 6.5 dB/m, with no other mode content observed. The same method was applied to Fiber #1 with a 20 m long length, showing a $LP_{11}$ mode loss of 0.43 dB/m. Note that even a HOM loss of 0.43 dB/m may be sufficient for optical communication applications [5,18]. For applications requiring very high mode purity, such as FOG, increasing the FM loss by just 0.03 dB/km (comparing Fiber #5 to Fiber #1) can increase the HOMER by more than an order of magnitude, from 4300 to 50,000.

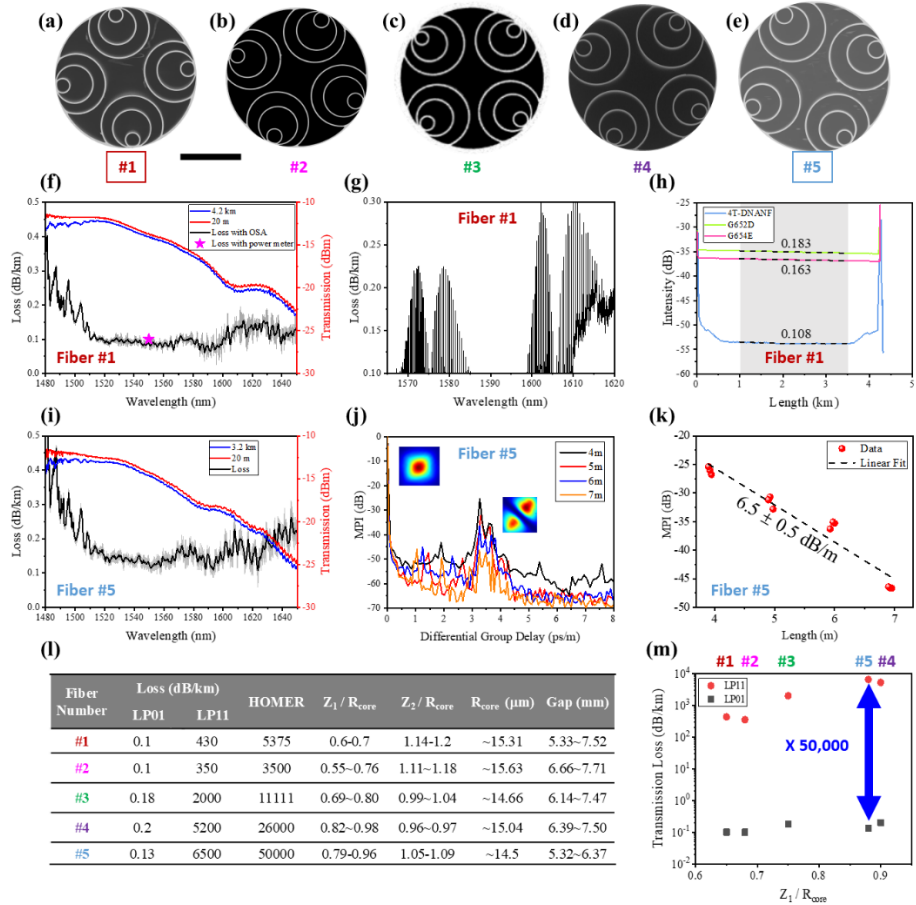

**Fig.2.** AR-HCF Fabrications and Loss Measurements. (a-e) SEM images of the five fabricated 4T-DNANFs with their structural parameters listed in (l). Scale bar, 50 μm. (f) Measured transmission (blue

and red) and loss (black) spectra of Fiber #1. The measurement uncertainty is indicated by the grey shaded area. (g) Loss spectrum of Fiber #1 measured with a wavelength resolution of 0.05 nm, revealing the absorption lines from $CO_2$ gas. (h) OTDR traces of a 4T-DNANF (Fiber #1), a G.654E fiber, and a G.652D fiber. (i) Same as (f) for Fiber #5. (j) $S^2$-imaging measurement of Fiber #5 with varying fiber lengths. The insets show the reconstructed mode-intensity profiles at two different differential group delays. (k) Fitted loss curve of the $LP_{11}$ mode derived from (j) for Fiber #5. (l) Summaries of structural parameters, modal losses, and HOMER for the five AR-HCFs. (m) Measured average losses of the $LP_{01}$ and $LP_{11}$ modes in the 1540-1560 nm range, along with the corresponding HOMER for Fibers #1 - #5.

The structural parameters and measured loss of $LP_{01}$ and $LP_{11}$ modes of Fibers #1 to #5 are listed in Fig. 2(l). All five fibers show FM losses under 0.2 dB/km. A clear trend of increasing $LP_{11}$ mode loss with $Z_1/R_{core}$ is observed, as shown in Fig. 2(m), confirming the importance of the first cladding air crescent in HOM filtering. These fibers demonstrate strategies for achieving ultralow FM loss with acceptable HOM loss or very high mode purity still with very low loss for various applications.

The fiber's other optical properties, such as chromatic dispersion (CD) and bending loss, are comparable to those other AR-HCFs with the same core size. The CD value is measured to be 2-4 ps/nm/km in the C-band, and no distinguishable bending loss is observed for radii of 8 cm over 300 turns. Therefore, these properties will not be elaborated upon here.

## 4. Discussion and Conclusions

Low loss AR-HCFs typically consist of multiple AR layers [14], including both silica membranes and air cavity. While the silica membranes are typically highlighted for their role in anti-resonant reflection, the air layers are equally important but often overlooked. These air layers serve two critical functions: they not only contribute to the anti-resonant guiding mechanism but also play a key role in stripping out higher-order modes (HOMs). The choice of a 4T-DNANF design over 5-DNANF is directly related to the optimization of the air cavity. By fine-tuning its dimensions, the 4T-DNANF achieves a superior balance between low FM loss and high HOM extinction, overcoming the inherent multimode nature of HCFs. The flexibility in tuning the core and cladding parameters also opens up new possibilities for further reducing loss and enhancing performance in future fiber designs. These findings mark a pivotal step forward in the development of next-generation optical fibers for advanced communication and sensing technologies.

## 5. Funding

This work was supported by the National Natural Science Foundation of China under Grants 62222506, 62105122, U21A20506, and 62075083, and in part by the Basic and Applied Basic Research Foundation of Guangdong Province 2021B1515020030, 2021A1515011646, and 2022A1515110218, and in part by the Guangzhou Science and Technology Program 2024A04J9899.

## 6. Acknowledgments

The authors acknowledge all the technicians for this work: Guohao Li, Xiaosong Lu, Zhiqing Li, and Luxu Li.

## 7. Data availability statement

Data underlying the results presented in this paper are not publicly available at this time but may be obtained from the authors upon reasonable request.